\documentstyle[12pt]{article}

\catcode`\@=11

\def\@citex[#1]#2{%
\if@filesw \immediate \write \@auxout {\string \citation {#2}}\fi
\@tempcntb\m@ne \let\@h@ld\relax \def\@citea{}%
\@cite{%
  \@for \@citeb:=#2\do {%
    \@ifundefined {b@\@citeb}%
      {\@h@ld\@citea\@tempcntb\m@ne{\bf ?}%
      \@warning {Citation `\@citeb ' on page \thepage \space undefined}}%
      {\@tempcnta\@tempcntb \advance\@tempcnta\@ne%
      \@tempcntb\number\csname b@\@citeb \endcsname \relax%
      \ifnum\@tempcnta=\@tempcntb %   Number follows previous--hold on to it
        \ifx\@h@ld\relax%
          \edef \@h@ld{\@citea\csname b@\@citeb\endcsname}%
        \else%
          \edef\@h@ld{\ifmmode{-}\else--\fi\csname b@\@citeb\endcsname}%
        \fi%
      \else%   %  non-successor--dump what's held and do this one
        \@h@ld\@citea\csname b@\@citeb \endcsname%
        \let\@h@ld\relax%
      \fi}%
    \def\@citea{,\penalty\@highpenalty\,}%
  }\@h@ld
}{#1}}
\def\@cite#1#2{{$^{#1}$\if@tempswa , #2\fi }}

\def\citel{\@ifnextchar [{\@tempswatrue\@citexl}{\@tempswafalse\@citexl[]}}
\def\@citexl[#1]#2{%
\if@filesw \immediate \write \@auxout {\string \citation {#2}}\fi
\@tempcntb\m@ne \let\@h@ld\relax \def\@citea{}%
\@citel{%
  \@for \@citeb:=#2\do {%
    \@ifundefined {b@\@citeb}%
      {\@h@ld\@citea\@tempcntb\m@ne{\bf ?}%
      \@warning {Citation `\@citeb ' on page \thepage \space undefined}}%
      {\@tempcnta\@tempcntb \advance\@tempcnta\@ne%
      \@tempcntb\number\csname b@\@citeb \endcsname \relax%
      \ifnum\@tempcnta=\@tempcntb %   Number follows previous--hold on to it
        \ifx\@h@ld\relax%
          \edef \@h@ld{\@citea\csname b@\@citeb\endcsname}%
        \else%
          \edef\@h@ld{\ifmmode{-}\else--\fi\csname b@\@citeb\endcsname}%
        \fi%
      \else%   %  non-successor--dump what's held and do this one
        \@h@ld\@citea\csname b@\@citeb \endcsname%
        \let\@h@ld\relax%
      \fi}%
    \def\@citea{,\penalty\@highpenalty\,}%
  }\@h@ld
}{#1}}
\def\@citel#1#2{{#1\if@tempswa , #2\fi}}

\newcount\hour
\newcount\minute
\newtoks\amorpm
\hour=\time\divide\hour by 60
\minute=\time{\multiply\hour by 60 \global\advance\minute by-\hour}
\edef\standardtime{{\ifnum\hour<12 \global\amorpm={am}%
        \else\global\amorpm={pm}\advance\hour by-12 \fi
        \ifnum\hour=0 \hour=12 \fi
        \number\hour:\ifnum\minute<10 0\fi\number\minute\the\amorpm}}
\edef\militarytime{\number\hour:\ifnum\minute<10 0\fi\number\minute}

\def\draftlabel#1{{\@bsphack\if@filesw {\let\thepage\relax
   \xdef\@gtempa{\write\@auxout{\string
      \newlabel{#1}{{\@currentlabel}{\thepage}}}}}\@gtempa
   \if@nobreak \ifvmode\nobreak\fi\fi\fi\@esphack}
        \gdef\@eqnlabel{#1}}
\def\@eqnlabel{}
\def\@vacuum{}
\def\marginnote#1{}
\def\draftmarginnote#1{\marginpar{\raggedright\scriptsize\tt#1}}

\def\draft{
	\pagestyle{plain}
        \oddsidemargin -.5truein
        \def\@oddhead{\sl \phantom{\today\quad\militarytime} \hfil
        \smash{\Large\sl DRAFT} \hfil \today\quad\militarytime}
        \let\@evenhead\@oddhead
        \let\label=\draftlabel
        \let\marginnote=\draftmarginnote
        \def\ps@empty{\let\@mkboth\@gobbletwo
        \def\@oddfoot{\hfil \smash{\Large\sl DRAFT} \hfil}
        \let\@evenfoot\@oddhead}
        \def\@eqnnum{(\theequation)\rlap{\kern\marginparsep\tt\@eqnlabel}%
        \global\let\@eqnlabel\@vacuum}  }

\def\blackfonts{
        \font\blackboard=msbm10 scaled\magstep1
        \font\blackboards=msbm8
        \font\blackboardss=msbm6
}

\def\nblack{            % For people without blackboard fonts
\def\ZZ{{Z \n{10} Z}}
\def\NN{{N \n{14} N}}
\def\CC{{C \n{11} C}}
\def\RR{{R \n{11} R}}
\def\QQ{{Q \n{12} Q}}
\def\PP{{P \n{11} P}}
}

\def\prep{         % twocolumn.sty  Changed my Marek and Neil
	\catcode`\@=11
        \input art10.sty
	\catcode`\@=12
	\let\small\null
	\def\blackfonts{
        	\font\blackboard=msbm10
        	\font\blackboards=msbm7
        	\font\blackboardss=msbm5
	}
	\let\sl\it
        \twocolumn
        \sloppy
        \voffset=-2.54truecm
        \hoffset=-2.54truecm
        \flushbottom
        \parindent 1em
        \leftmargini 2em
        \leftmarginv .5em
        \leftmarginvi .5em
        \marginparwidth 48pt
        \marginparsep 10pt
        \setlength{\columnsep}{2truecm}
        \setlength{\textwidth}{25.4truecm}
        \setlength{\textheight}{17truecm}
        \baselineskip=16pt
        \oddsidemargin .18truein
        \evensidemargin .17truein
}

\def\eqalign#1{\null\,\vcenter{\openup\jot\m@th
  \ialign{\strut\hfil$\displaystyle{##}$&$\displaystyle{{}##}$\hfil
      \crcr#1\crcr}}\,}
\def\eqalignno#1{\displ@y \tabskip\centering
  \halign to\displaywidth{\hfil$\@lign\displaystyle{##}$\tabskip\z@skip
    &$\@lign\displaystyle{{}##}$\hfil\tabskip\centering
    &\llap{$\@lign##$}\tabskip\z@skip\crcr
    #1\crcr}}

\def\section{\@startsection {section}{1}{\z@}{3.ex plus 1ex minus
 .2ex}{2.ex plus .2ex}{\large\bf}}
\def\subsection{\@startsection{subsection}{2}{\z@}{2.75ex plus 1ex minus
 .2ex}{1.5ex plus .2ex}{\bf}}

\def\appendix{\let\appendix\moreappendix%
        \setcounter{section}{0} \setcounter{subsection}{0}
        \@addtoreset{theorem}{section}
        \def\thesection{\Alph{section}}\moreappendix}
\def\moreappendix{\@startsection {section}{1}{\z@}{3.ex plus 1ex minus
 .2ex}{2.ex plus .2ex}{\large\bf Appendix }}
\def\thefootnote{\fnsymbol{footnote}}
\def\abstract{\if@twocolumn
\section*{Abstract}
\else %\small
\begin{center}
{\bf Abstract\vspace{-.5em}\vspace{0pt}}
\end{center}
\quotation
\fi}

\catcode`\@=12

\def\adhoc{{\it ad hoc\/}}

\def\ansatz{{\it ansatz\/}}

\def\Mobius{M\"obius}
\def\Kahler{K\"ahler}
\def\kahler{K\"ahler}
\def\apriori{{\it a priori\/}}

\def\etal{{\it et al.\/}}
\def\etc{{\it etc.\/}}
\def\ala{{\it \`a la\/}}

\def\ibid{{\it ibid.\/}}

\def\noj#1,#2,{{\bf #1} (19#2)\ }
\def\jou#1,#2,#3,{{\sl #1\/ }{\bf #2} (19#3)\ }
\def\ann#1,#2,{{\sl Ann.\ Physics\/ }{\bf #1} (19#2)\ }
\def\cmp#1,#2,{{\sl Comm.\ Math.\ Phys.\/ }{\bf #1} (19#2)\ }
\def\cq#1,#2,{{\sl Class.\ Quantum Grav.\/ }{\bf #1} (19#2)\ }
\def\cqg#1,#2,{{\sl Class.\ Quantum Grav.\/ }{\bf #1} (19#2)\ }
\def\ijmp#1,#2,{{\sl Int.\ J.\ Mod.\ Phys.\/ }{\bf A#1} (19#2)\ }
\def\jmp#1,#2,{{\sl J.\ Math.\ Phys.\/ }{\bf #1} (19#2)\ }
\def\grg#1,#2,{{\sl Gen.\ Rel.\ Grav.\/ }{\bf #1} (19#2)\ }
\def\mpl#1,#2,{{\sl Mod.\ Phys.\ Lett.\/ }{\bf A#1} (19#2)\ }
\def\nc#1,#2,{{\sl Nuovo Cim.\/ }{\bf #1} (19#2)\ }
\def\np#1,#2,{{\sl Nucl.\ Phys.\/ }{\bf B#1} (19#2)\ }
\def\pl#1,#2,{{\sl Phys.\ Lett.\/ }{\bf #1B} (19#2)\ }
\def\pla#1,#2,{{\sl Phys.\ Lett.\/ }{\bf #1A} (19#2)\ }
\def\pr#1,#2,{{\sl Phys.\ Rev.\/ }{\bf #1} (19#2)\ }
\def\prd#1,#2,{{\sl Phys.\ Rev.\/ }{\bf D#1} (19#2)\ }
\def\prl#1,#2,{{\sl Phys.\ Rev.\ Lett.\/ }{\bf #1} (19#2)\ }
\def\prp#1,#2,{{\sl Phys.\ Rept.\/ }{\bf #1C} (19#2)\ }
\def\ptp#1,#2,{{\sl Prog.\ Theor.\ Phys.\/ }{\bf #1} (19#2)\ }
\def\ptpsup#1,#2,{{\sl Prog.\ Theor.\ Phys.\/ Suppl.\/ }{\bf #1} (19#2)\ }
\def\rmp#1,#2,{{\sl Rev.\ Mod.\ Phys.\/ }{\bf #1} (19#2)\ }
\def\yadfiz#1,#2,#3[#4,#5]{{\sl Yad.\ Fiz.\/ }{\bf #1} (19#2) #3%
\ [{\sl Sov.\ J.\ Nucl.\ Phys.\/ }{\bf #4} (19#2) #5]}
\def\zh#1,#2,#3[#4,#5]{{\sl Zh.\ Exp.\ Theor.\ Fiz.\/ }{\bf #1} (19#2) #3%
\ [{\sl Sov.\ Phys.\ JETP\/ }{\bf #4} (19#2) #5]}

\def\eq#1{.~(\ref{#1})}
\def\beq{\begin{equation}}
\def\eeq{\end{equation}}
\def\beqar{\begin{eqnarray}}
\def\eeqar{\end{eqnarray}}

\def\nfrac#1#2{{\displaystyle{\vphantom1\smash{\lower.5ex\hbox{\small$#1$}}%
        \over\vphantom1\smash{\raise.25ex\hbox{\small$#2$}}}}}

\def\p#1{\mskip#1mu}
\def\n#1{\mskip-#1mu}
\def\stop{\p6.}
\def\comma{\p6,}

\def\eqand{\p6 {\rm and}}

\def\to{\rightarrow}

\def\longlongrightarrow{\relbar\joinrel\relbar\joinrel\rightarrow}

\def\onArrow#1{\mathrel{\mathop{\longlongrightarrow}\limits^{#1}}}

\def\lae{\mathrel{\mathop{\smash{\lower .5 ex \hbox{$\stackrel<\sim$}}}}}
\def\lae{\mathrel{\mathop{\smash{\lower .5 ex \hbox{$\stackrel>\sim$}}}}}

\def\vev#1{\left\langle #1 \right\rangle}

\def\f{\frac}
\def\pa{\partial}
\def\pb{\bar\pa}

\def\Tr{{\rm Tr}}
\def\l:{\mathopen{:}\,}
\def\r:{\,\mathclose{:}}

\def\nq{$N=2$}
\def\twot{$(2,2)$}
\def\xb{{\bar{x}}}
\def\kb{{\bar{k}}}
\def\jb{{\bar{j}}}
\def\ib{{\bar{i}}}

\def\zb{{\bar{z}}}
\def\Xb{{\bar{X}}}
\def\kb{{\bar{k}}}
\def\pb{{\bar{\pa}}}
\def\tb{{\bar{\theta}}}
\def\psb{{\bar{\psi}}}

\def\dx{d x}
\def\dxf{d^{\p1 4}\n2 x}
\def\dxt{d^{\p1 2}\n2 x}
\def\dzt{d^{\p1 2}\n2 z}
\def\dtt{d^{\p1 2}\theta}

\def\dttb{d^{\p1 2}\tb}

\def\sut{SU(2)}

\def\son{SO(N)}
\def\sot{SO(2)}
\def\spn{USp(N)}
\def\un{U(N)}
\def\sott{SO(2,2)}
\def\uo{U(1)}
\def\uoo{U(1,1)}

\def\rptwo{\RR P^2}

\def\sklein{\hbox{\scriptsize\sl Klein}}
\def\sannulus{\hbox{\scriptsize\sl annulus}}
\def\storus{\hbox{\scriptsize\sl torus}}
\def\smobius{\hbox{\scriptsize\sl M\"obius}}

\def\beq{\begin{equation}}
\def\eeq{\end{equation}}
\def\eq#1{.~(\ref{#1})}

\nblack
\begin{document}

\parindent=1.5pc
\parskip=6pt
\def\thefootnote{\fnsymbol{footnote}}

\begin{titlepage}

\noindent November 13, 1992.		\hfill    TAUP--2002--92 \\
\null\hfill    hep-th/9211059

\vskip 1 cm

\begin{center}{{\bf
A tour through $N=2$ strings.
               }\\
\vglue 1.0cm
{NEIL MARCUS\footnote{
Work supported in part by the US-Israel Binational Science Foundation,
and the Israel Academy of Science.  E-Mail: NEIL@HALO.TAU.AC.IL.}
}\\
\baselineskip=14pt
{\it School of Physics and Astronomy\\Raymond and Beverly Sackler Faculty
of Exact Sciences\\Tel-Aviv University}\\
\baselineskip=14pt
{\it Ramat Aviv, Tel-Aviv 69978, ISRAEL}\\
% \vglue 0.3cm
% {and}\\
% \vglue 0.3cm
% {SECOND AUTHOR'S NAME}\\
% {\it Group, Company, Address, City, State ZIP/Zone, Country}\\
%\vglue 0.8cm
%{ ABSTRACT }
}
\end{center}
%\vglue 0.3cm
%{\rightskip=3pc
 %\leftskip=3pc
 %\tenrm\baselineskip=12pt
% \noindent

\begin{abstract}

I give an overview of open, closed and heterotic \nq\
strings.  At the tree level I derive the effective field theories of all
the strings, and discuss the group theory of the \nq\ open string and
the interaction between its open and closed sectors.  The
two-dimensional effective field theory of the open \nq\
string is a sigma model, while the four-dimensional theory gives
self-dual Yang-Mills (SDYM) in a self-dual gravity (SDG) background.
The theory can have any gauge group, unlike the usual Chan-Paton
\ansatz.  The four-dimensional closed string gives SDG, and the
heterotic string is related to SDYM.  At one loop \nq\
string loop amplitudes and partition functions have incurable infra-red
divergences, and show puzzling disagreements on the dimension of
spacetime when compared to their effective field theories.  I show that
the known closed-string three-point amplitude can be written directly
in terms of a Schwinger parameter, so explicitly exhibiting the
inconsistency.  I finally discuss the possibility that the puzzles
posed by the loop amplitudes could be solved if the \nq\
theories were Lorentz invariant and supersymmetric, and I speculate on
possible modifications of the string calculations.

\vskip 1cm
{\rightskip=3pc
 \leftskip=3pc
Talk given
at the International workshop on ``String theory, quantum gravity and
the unification of fundamental interactions'', Rome 1992.}

\end{abstract}
%}
\newpage
\end{titlepage}
\setcounter{footnote}{0}

\def\twoc{$2_\CC$}
\def\twor{$2_\RR$}
\def\fourr{$4_\RR$}
\def\ad{Ademollo \etal}

\baselineskip=14pt

{\bf\noindent 1. The basics of \nq\ strings}
\vglue 0.2cm
{\it\noindent 1.1. Introduction}
\vglue 0.1cm

I would like to thank the organizers of this conference for giving me the
opportunity to talk about \nq\ strings in Rome.  It would be a pleasure to
give any talk here, but it is particularly appropriate to talk about \nq\
strings, since they were first studied in the famous papers by
\ad\cite{ademonly,adem}, and so have 11 Italian (and 2 foreign) fathers.
Now, although \nq\ strings naturally live in four dimensions, it is a
four-dimensional spacetime with signature \twot\ or $(4,0)$.  Thus, unlike
the subjects of the first two talks here, \nq\ strings can not be directly
relevant to nature.  So, before jumping into the physics, I would
like to give you some reasons for talking to you about them, aside from
the aptness of the location.

First, there are interesting
mathematical aspects to these theories.  As we shall see, the
field theories describing these strings are self-dual gravity (SDG) for the
closed string and self-dual Yang-Mills (SDYM) for the open and heterotic
strings.  SDYM in Euclidean space is of course related to instantons,
which have a whole host of interesting physical and mathematical
properties. In addition, the dimensional reductions of SDYM to two
dimensions gives rise to many integrable systems\cite{integ2}, and it has
even been conjectured that all two-dimensional integrable systems are
reductions of SDYM\cite{conject}.  \nq\ strings thus may give us some notion of
what a quantum theory of instantons should be.

What is more interesting from my point of view, and perhaps also from the
point of view of the fathers of the string, are the ``stringy'' reasons
for examining these theories.  Although \nq\ strings are unrealistic,
studying them may provide us with better insight into string theory in
general because, in some ways, they are remarkably simple theories.
An example of this will be our ability to directly compare the loop amplitudes
of the \nq\ strings to those of their effective field theories.  Since \nq\
strings do not have Regge trajectories, and contain only massless
particles, they give rise to field theories containing only a finite
number of fields.  In this way they are similar to the $c=1$ theories, but
are even more simple since, being Lorentz invariant, they can not contain
any discrete states.   Also, \nq\ amplitudes are local and many of them
vanish\cite{OV,OVh,me}, possibly indicating some kind of topological
structure of the theories.  It will turn out that, despite their
simplicity, our understanding of \nq\ theories is still immature with
many puzzles remaining.  I would like to view this as a
challenge, rather than a problem, and I hope that other people will be
encouraged to enter into the subject.

The original work that I will present in this talk  consists of the
reconsideration of the \nq\ open string in \twoc\ dimensions\cite{me}.
However, instead of concentrating on this particular case, I shall rather
give a general tour through various \nq\ strings, telling you their status
and pointing out their problems.  This means that I shall be discussing
the work of many other people, with additional commentary of my own in
various places, and I shall of course be biased to those aspects of the
theories that are particularly
interesting to me.  For example, I shall not discuss
non-critical \nq\ strings at all, nor the \nq\ strings with background
charges\cite{noncrit}.  Also I shall not go into too much detail
on aspects of any particular theory; you can find more details in the
various original references.

The outline of my talk is the follows:
I first give a basic review of \nq\
strings.  Then I calculate various tree-level string amplitudes,
to find the effective field theory actions describing the strings.  I
then present some one-loop amplitudes to illustrate the problems
intrinsic in the amplitudes, and in comparing the amplitudes to those of
the effective field theory.  I conclude with a discussion of the status
of the \nq\ strings, and of possible solutions to their various problems.

\vskip 0.2cm
{\it\noindent 1.2. \nq\ supergravity in two dimensions}
\vglue 0.1cm

At this stage, I should explain to those who do not know that the ``$N$''
of \nq\ refers to the number
of local world-sheet supersymmetries of the string.
Thus the $N=0$
string is the bosonic string, which is basically a theory of matter
coupled to 2-dimensional world-sheet gravity, and the $N=1$ string is a
theory of supermatter coupled to world-sheet $N=1$ supergravity.  (In heterotic
strings, the numbers of left- and right-handed supersymmetries are
different.)  \nq\ supergravity was first considered by Brink and
Schwarz\cite{BrinkS}.
The \nq\ supergravity multiplet consists of a
vielbein, a {\it complex\/} gravitino and a
$\uo$ gauge field denoted, respectively:
\beq
e_\alpha^a \comma \quad
\left(\begin{array}{c} \chi_\alpha^{(1)} \\
                       \chi_\alpha^{*\p1(-1)} \end{array}\right)
\quad {\rm and} \quad
A_\alpha \stop
\eeq
Here the numbers in parenthesis show the $\uo$ charges of the gravitini.
As is usual in string theory, all the supergravity fields can
be locally gauged away using the gauge symmetries of the
theory, and there is no action for them:  The vielbein is removed by
general coordinate invariance, local Lorentz invariance and local Weyl
transformations; the gravitini by the complex supersymmetry and complex
super-Weyl transformations, and the two components of the $\uo$ gauge field
by vector and chiral $\uo$ gauge symmetries on the world sheet\footnote{The
existence of
the chiral invariance was pointed out by Fradkin and Tseytlin
in ref.~\citel{Frad}.}.  These $\uo$ symmetries have no counterpart in the
$N=0$ and $N=1$ theories, and their existence leads to many of the
special features
of \nq\ strings.

To get a string theory, one must couple the supergravity to some (\nq)
matter.  The simplest such matter is an \nq\ chiral superfield $X^i$, with
$i$ some internal (spacetime!) index.  The component fields are seen in
the $\theta$ expansion of these superfields given, somewhat schematically,
by:
\beq
X^i \sim x^i + \theta^{(-1)} \psi^{i \p1 (1)}
\quad \eqand \quad
\Xb^\ib \sim \xb^\ib + \theta^{*\p1 (1)} \psi^{* \p1 \ib \p1 (-1)}
\stop
\label{fields}
\eeq
The string world-sheet action is\cite{BrinkS}:
\beq
\eqalign{
S = \int\dzt \sqrt{g} \, \Bigl( \,
&\nfrac12 \, g^{\alpha\beta} \pa_\alpha x_i \pa_\beta \xb^\ib
   + i \, \bar{\psi}^\ib  D \n{11} / \p{5} \psi_i
   + A_\alpha \bar{\psi}^\ib \gamma^\alpha \psi_i \cr
& + (\pa_\alpha \xb^\ib + \bar\psi^\ib \chi ) \, \chi_\beta \gamma^\alpha
\gamma^\beta \psi_i \;+\:\hbox{c.c.} \,  \Bigr)
\label{wsa} \comma
}
\eeq
where $D \n{11} / \p{5} $ denotes a gravitationally covariant derivative
of the spinor, containing a spin-connection piece.
Note that, although the scalar field $x^i$ is complex, it does not
have a
$\uo$ charge and does not couple to the gauge field $A_\alpha$.
On the other hand, the
spinors $\psi^i$ are charged, and I have explicitly shown
their minimal coupling in the action.

\vskip 0.2cm
{\it\noindent 1.3. The critical dimension of \nq\ strings}
\vglue 0.1cm

The first question to ask in any string theory is what is the
critical dimension of the theory?  \ad\ found that in \nq\ strings
$D=2$, where $D$ is the number of chiral superfields
$X^i$.  The modern derivation of this result goes as follows\cite{Frad}:
gauge fixing the $N=2$ supergravity algebra gives rise to independent
left and right-handed \nq\ constraint algebras:
\beq
\left(\begin{array}{c} T \\
                       G \quad G^* \\
                       J \end{array}\right)_L \quad {\rm and} \quad
\left(\begin{array}{c} \bar T \\
                       \bar G \quad \bar {G^*} \\
                       \bar J \end{array}\right)_R \stop
\label{constr}
\eeq
In both the left and right sectors
one has
the usual $(b,c)$ ghosts, {\it complex\/}
$(\beta,\gamma)$ supersymmetry ghosts and
a $(b',c')$ system of
$\uo$ ghosts with conformal spins $(1,0)$.   Canceling the conformal anomaly
means
\beq
c =-26+2\cdot11-2+2D \cdot (1+\nfrac12) = 0 \comma
\eeq
resulting in $D=2$.  An even easier calculation is to cancel the axial-vector
$\uo$ anomaly between the $\psi$'s and the $(\beta,\gamma)$ ghosts:
\beq
c=0=-2+D \stop
\eeq
This gives the same result, since all the anomalies of the theory fall into a
single supermultiplet.

Since we have 2 {\it complex\/} scalars $x^i$, it would now
appear to be obvious
that the theory should live in a \twoc-spacetime; in terms of real
dimensions this is a \fourr-dimensional spacetime that must have a
signature $(2,2)$ or $(4,0)$.  However, here the history of the subject is
a little peculiar.  In their original work, \ad\ defined a real superfield
$Y = X + \Xb$, instead of working with the chiral superfield $X$.
This superfield contains the same information as $X$,
except that the imaginary part of the zero-mode $x$ appears only as a
derivative. Based on this, they decided not to excite positions and
momenta in these ``imaginary'' coordinates.  This means that
they dimensionally reduced
the theory by brute force to \twor\ dimensions.  (Note that the anomaly
cancellation still works in the truncated theory, since only the zero-modes
of the fields have been
changed, and the theory still contains 4 real bosonic and fermionic
fields.)  When Fradkin and Tseytlin rederived the critical
dimension of the theory they considered the imaginary coordinates to be
physical\cite{Frad}, and this was later taken by D'Adda and Lizzi
to imply a Lorentz-invariant \fourr-dimensional spacetime\cite{4d}.
Since then the theory has been firmly ensconced in \twoc\ dimensions.

\vskip 0.2cm
{\it\noindent 1.4. The spectrum of \nq\ strings}
\vglue 0.1cm

To see the spectrum of the theory one needs to calculate the theory on the
cylinder.  This means that we must discuss the boundary conditions of the
fields. As usual, since they are world-sheet tensor fields, the $x^i$
fields and the anticommuting ghosts are periodic (although other possibilities
were considered in ref.~\citel{adem}).  However, a new feature of the \nq\
string is that the fermions $\psi^i$ and the commuting ghosts are charged
under the $\uo$ symmetry.  Since one can have a constant gauge field on
the cylinder (more precisely one can have non-trivial Wilson lines around
the cylinder), one can continuously change the boundary conditions of the
fermions from NS to R by turning on this gauge field.  Because of this,
one usually considers just the purely NS part of the theory, and argues
that all other sectors are equivalent to it by the \nq\ spectral
flow\cite{spect}.  I shall generally do this throughout the talk, although
I shall discuss other possibilities in the conclusion.

Now naively, since we are in $D=2$, one would expect that one could go to
a light-cone gauge with $D-2=0$ transverse dimensions!  This suggests that
\nq\ theories should have no oscillator excitations.  The light-cone
argument also suggests that the mass squared of this state is $(2-D)/24=0.$
The decoupling of the massive states of the string
was indeed seen by \ad, and the same
result has been confirmed by a BRST analysis\cite{BRST}.  But the simplest way
to find the spectrum
is to calculate the partition function of the theory: Aside
from zero modes, the four $x$ oscillators are canceled by the $(b,c)$ and
$(b',c')$ ghosts, while, for any boundary conditions, the two complex charged
$\psi^i$'s are canceled by the charged $(\beta,\gamma)$ ghost system.
This leaves only massless states in the spectrum.
Summarizing:
\begin{itemize}
{\it
\item[a)] \nq\ strings live in $2_\CC$ dimensions.
\item[b)] All their oscillator excitations vanish.
\item[c)] In each sector of the theory, only a massless ``scalar'' state
          propagates.
}
\end{itemize}
\noindent and, thus:
\begin{itemize}
{\it
\item[d)] \nq\ string amplitudes must satisfy duality with no infinite sums
          over massive states!
\item[e)] \nq\ string field theory is an ordinary field theory with a
finite number of particles.
}
\end{itemize}

\vskip 0.2cm
{\it\noindent 1.5. The Lorentz symmetry of the \nq\ strings}
\vglue 0.1cm

We have now seen where the \nq\ strings live and have calculated their
spectrum. One remaining question that we should ask is
what is the ``Lorentz'' invariance of these theories.  Note that in the
gauge-fixed theory, where one
turns off the supergravity fields $A_\alpha$ and $\chi_\alpha$ in the
string action of eq\eq{wsa}, there is no difference between matter fields
and their complex conjugates in the action.  Based on this, D'Adda and
Lizzi argued that the the theory has an $\sott$ Lorentz invariance\cite{4d}
(chosing the
``Minkowski'' rather than the Euclidean signature).  However, the
$\psi^i${}'s and $\psi^{* \p1 \ib}${}'s couple differently to the
gauge field $A_\alpha$, since they have
opposite charges. This difference is then fed to the $x^i$ and
$\xb^\ib$ fields by
the gravitini $\chi_\alpha$.  Another way of looking at this is that
the constraint multiplets of
eq\eq{constr} distinguish the different fields.  In the works of Ooguri
and Vafa that sparked the renewed interest in \nq\ strings\cite{OV},
spacetime is
thus considered to be intrinsically complex (actually \kahler\/), with a
$\uoo$ symmetry group.

Recently, however, Siegel has
argued\cite{Warren} that the \nq\ string is the same as the ``$N=4$''
string, which was also discovered by \ad\cite{adem}.  This
string has an $\sut$ local symmetry, is hard to
quantize covariantly, and apparently has $D=-2$!  If
Siegel is correct, the manifest $\uoo$ invariance of the \nq\
string should be extendible to the full $\sott$ Lorentz invariance.  This
is known to be the case in the tree-level amplitudes of
refs.~\citel{OV,OVh,me}, as I will discuss. A bigger Lorentz
symmetry would be important, since the $\uoo$
spaces have no little group,
so one cannot define the ``spins'' of particles, while the $\sott$
Poincar\'e group
can have representations of continuous spin.  Using the larger Lorentz
group, Siegel has also argued that
the Ramond sectors of the \nq\ strings should describe fermions\cite{Warren},
and that the effective field theories of the \nq\ strings
should be maximally spacetime supersymmetric\cite{Warren2}.  Such results
necessitate
a modification of the string amplitudes. I shall return to these issues in the
discussion section at the end of my talk.

\vskip 0.4cm
{\bf\noindent 2. Tree amplitudes and effective actions}
\vglue 0.2cm
{\it\noindent 2.1. \nq\ strings in two {\it real\/} dimensions}
\vglue 0.1cm

We have seen that \nq\ strings do not have any oscillator excitations, and
so can be described by effective field theories with finite numbers of
fields.  The most straightforward approach to finding
these field
theories is simply to calculate the tree-level amplitudes of the
string theory, and to equate them to those of the field theory.
Here I shall follow an approximately historical approach, and first
consider the
strings in two {\it real\/} dimensions, \ala\ \ad{}
It may seem a bit strange to the
post-Polyakov physicist that the \twor-dimensional theory makes any
sense at all.   However, we have already argued that (local) anomalies are
still canceled in this truncated theory, and tree amplitudes are
consistent even in dimensions which do not give anomaly cancellation. We
would expect this truncation to make troubles at some stage, but since the
rules for writing down \nq\ strings are so ill-defined, it is even
possible that these theories are consistent as is.  It
turns out that these truncated theories illustrate many of the general
properties of \nq\ strings, and have
interesting effective field theories in their own right.

I now turn to the case of the open string in \twor\
dimensions\cite{adem}.  As I
suggested above, the spectrum of this theory should be a single massless
scalar.
However, as is usual in open strings, \ad\ added ($\sut$) Chan-Paton
factors\cite{CP} to the string, ending up with a theory
of three scalars in the adjoint
of \sut.  Since the \twor-dimensional theory is the truncation of the
\twoc\ theory, which I shall consider in detail later, I shall simply
present its amplitudes without derivation:  The first
nonvanishing amplitude in the theory is the 4-point function, since all
odd-point functions vanish by the G-parity of the
theory.  It is given by
\beq
A_4 = \frac{g^2}2 \, u \; \Tr \left( \lambda_1 \lambda_2 \lambda_3 \lambda_4
        \right) + {\hbox{perms}} \p6,
\label{adem4}
\eeq
where $u$ is the usual Mandelstam variable, and the $\lambda$'s
are the Chan-Paton factors.  Note that this amplitude would vanish
without the Chan-Paton factors, since
it would be proportional to $s+t+u=0$.  Also, recall that in
the scattering of two massless particles in \twor-dimensions one has the
kinematic relation
\beq
stu=0 \stop
\label{stu}
\eeq
One can get the equation of motion corresponding to
eq\eq{adem4} from an $\sut$ sigma model\footnote{Because
the sign of the
amplitude has not been fixed, the correct theory may be
the coset
theory $\sut_\CC/\sut$\cite{OVh}.}, with action
\beq
S= \int \dxt \, \Tr \left( g^{-1} \pa_i g \; g^{-1} \pa^i g \right) \comma
\eeq
and equation of motion
\beq
\pa_i \left( g^{-1} \pa^i g\right) =0 \stop
\label{ademeq}
\eeq
\ad\ argued that this equivalence should be true to all orders, using
uniqueness arguments based on Regge behaviour and Adler zeroes, which I do
not claim to follow.

We can already see several intriguing features of this theory, which will
generalize to all \nq\ theories.  Unlike
other string theories, and about any other field theories,
the amplitude of eq\eq{adem4} is a local function of the
momenta.  That this had to happen is pretty much forced on us by the
opposing requirements of the duality of the amplitudes and the fact that
the theory contains only massless particles.  This should therefore
be true for all nonvanishing amplitudes.  \ad\
argued that
all $4N+2$ point functions vanish.  (Thus the next possibly nonvanishing
amplitude is the eight-point function---not surprisingly, it has never
been calculated.)  It will turn out that there are many
vanishing amplitudes in \nq\ string calculations, for reasons that are not
completely clear.  Finally, as one might hope for a theory based on
strings, the effective field theory has a nice geometrical nature.

I shall now briefly turn to the case of closed and heterotic strings in
\twor\ dimensions.  These were both studied in a paper by Green in
1987\cite{Mike}.  Using the rules for obtaining closed-string
amplitudes from open ones\cite{openclosed}, he found that the
closed \twor-dimensional \nq\ string has a vanishing four-point function, and
is presumably free.  On the other hand, the heterotic string has a
four-point amplitude given by
\beq
A_4^{\rm het} =
A_4^{\rm open} + \alpha' \, t u \, \delta_{12} \delta_{34} \comma
\eeq
which can be interpreted as a sigma-model amplitude with a correction from
a graph with an internal ``graviton''.

\vskip 0.2cm
{\it\noindent 2.2. Closed \nq\ strings in \twoc-dimensions}
\vglue 0.1cm

Continuing on the historical path, I now turn to a discussion of the
closed \nq\ string in \twoc\ dimensions, following Ooguri and
Vafa\cite{OV}.  I will consider this case in somewhat more detail, since I
will take over many of the conventions and results for the open case.
The world-sheet action of the theory, with all the supergravity fields
gauge fixed, can be written as
\beq
S = \int \frac{\dzt}\pi \, \dtt \, \dttb \, K_0(X^i, \Xb^\ib) \comma
\label{wsaf}
\eeq
where $X^i$ is the \nq{} chiral superfield.  Here $i$ runs from
0 to 1, corresponding to a
real (2,2)-dimensional spacetime, and
$K_0$ is the flat \Kahler\ potential $\eta_{i\jb}X^i \Xb^\jb$.
The only state of the theory is a
single massless scalar, corresponding to a perturbation around the flat
\kahler\ potential.  The
superspace-vertex operator for
emitting this scalar with (complex) momentum $k$ is
\beq
V_c = \f\kappa\pi \, e^{i(k \cdot \Xb + \kb \cdot X )} \stop
\eeq

The first amplitude that one can calculate is the three-point function.
Fixing the super-M\"obius transformations, this is given by
\beq
\eqalign{
A_{ccc} &= \vev{V_c|_{\theta=0}(0) \cdot \int \dtt\dttb \, V_c(1)
\cdot V_c|_{\theta=0}(\infty)}  \cr
&= \vphantom{\int} \kappa \, c_{12}^2 \comma
\label{a3c}
}
\eeq
where
\beq
c_{12} \equiv ( k_1 \cdot \kb_2 -\kb_1 \cdot k_2 )
\label{c12}
\eeq
is the extra invariant product of the momenta (other than the usual dot
product which has a plus sign) that exists in \twoc\ dimensions, when the
Lorentz group $\sott$ is reduced to $\uoo$.  Note that $c_{ij}$
is antisymmetric with respect to its two indices, and is additive in
the sense that $c_{i,j}+c_{i,k} = c_{i,j+k}$.  Using momentum
conservation, one sees that $A_{ccc}$ is totally symmetric, as it should
be.
It is important to note that in \twot\ dimensions, unlike the familiar
$(3,1)$-dimensional case, there is sufficient phase space to describe
one massless particle splitting into two others.  The
three-point function of the \twoc-dimensional closed string therefore
implies a truly nontrivial
S-matrix element, and not just some unphysical vertex.

The four-point function can be
calculated similarly, and results in an apparently
standard string amplitude:
\beq
\eqalign{
A_{cccc} &\sim \int \dzt \vev{V_c|_{\theta=0}(0) \cdot
\int \dtt\dttb \, V_c(z) \cdot \int \dtt\dttb \, V_c(1)
\cdot V_c|_{\theta=0}(\infty)}  \cr
&= \vphantom{\int} \f{\kappa^2}{\pi}
         F^2 \, \f { \Gamma(1-s/2) \, \Gamma(1-t/2) \, \Gamma(1-u/2) }
                 { \Gamma(s/2) \, \Gamma(t/2) \, \Gamma(u/2) } \stop
\label{a4c}
}
\eeq
However, Ooguri and Vafa noted that, like the \twor-dimensional
kinematic identity \hbox{$stu=0$},
in the scattering of massless particles
in \twot{}
dimensions there is again a relation:
\beq
F \,\equiv\, 1 - \f{c_{12}c_{34}}{su} - \f{c_{23}c_{41}}{tu} = 0 \stop
\label{F}
\eeq
This means that the four-point function of eq\eq{a4c} vanishes on
shell.  Ooguri and Vafa proposed that this vanishing comes from
some kind of topological nature of the theory, but this
is still somewhat of a mystery.
They also conjectured that all higher-point functions
in the \twoc-dimensional case
vanish; this is, as yet, unchecked.
(As I argued before, duality arguments just show that the amplitudes
should be local.)
It is simple to see that if the string is reduced to
\twor\ dimensions, as in the work of Green\cite{Mike}, the theory becomes
trivial:   In \twor\ dimensions $c_{ij}$
vanishes identically, so the three-point function now vanishes.  Also,
while in \twor\ dimensions $F \to 1$, the four-point function still vanishes
because of the identity $stu=0$.

Returning to \twoc\ dimensions, one sees that the local three-point function
and
vanishing four-point function can be obtained from
the action\cite{OV}
\beq
{\cal L}_c = \int \dxf \, \left( \nfrac12 \, \pa^i \phi \, \pb_\ib \phi +
\f{2\kappa}3 \, \phi \, \pa \pb \phi \wedge \pa \pb \phi + O(\phi^5) \right)
\comma
\label{pleb}
\eeq
where the $O(\phi^5)$ term needs to be determined from the five-point
function.  If all higher-order terms are absent the equation of motion of
this action has a name---the Plebanski equation\cite{Plebanski}---and has a
nice geometrical meaning.  It is therefore very reasonable that the action
is exact.  (Note that this conjecture and the conjecture that higher-point
amplitudes vanish are logically independent.) The Plebanski equation is
\beq
\pa^i \pb_\ib \phi - 2 \kappa
\, \pa \pb \phi \wedge \pa \pb \phi =0
\stop \label{plebeq}
\eeq
Its geometrical meaning is seen by considering $\phi$ to be
a perturbation of a \kahler\ potential around the flat \twoc\ space,
resulting in a spacetime metric
\beq
g_{i\jb} = \pa_i \pb_\jb \left( x_k \xb^\kb + 4 \kappa \phi \right) \stop
\label{nk}
\eeq
The Plebanski equation of motion then becomes simply
\beq
\det g_{i\jb} = -1 \stop
\eeq
Now since in a \kahler\ space the Ricci tensor $R_{i\jb}$ is given by
\beq
R_{i\jb} = \pa_i \pb_\jb \log \det g \comma
\label{ricci}
\eeq
the Plebanski equation is the condition for the Ricci flatness
of the $2_\CC$-dimen\-sional \kahler\ space.  An important theorem of Atiyah,
Hitchin and Singer\cite{math}, which is
quite nontrivial in the forward direction,
is that a \fourr-dimensional Riemann
space is self dual {\sl iff\/} it is \kahler\ and Ricci flat.  Thus the
equation of motion of the \nq\ closed string can be written elegantly as
\beq
R = \tilde R \comma
\eeq
and the \nq\ string describes self-dual gravity.
It is intriguing that although the entire formulation of the string has
been carried out
in a \twoc-dimensional \kahler\ space, the final equation of motion
can be written in a completely \twot\ Lorentz-invariant way.  I shall
discuss this further at the end of the talk.

\vskip 0.2cm
{\it\noindent 2.3. Purely open \nq\ strings in \twoc-dimensions}
\vglue 0.1cm

Since, despite apparently being a scalar theory, the closed \nq\ string
turns out to describe self-dual gravity,
it is reasonable to expect that the open and heterotic strings
describe self-dual Yang-Mills.  In the heterotic string\cite{OVh}, which
was studied by Ooguri and Vafa soon after the closed case, this is
basically true, but there are some complications.  Since the
geometrical structure of the heterotic string
is not that well understood, I shall leave the historical path and turn
rather to the open string, returning to make a few
comments on the heterotic string later.

I shall thus now turn to my work; the reconsideration of
the open string in \twoc\ dimensions\cite{me}\footnote{There were
previous attempts to calculate open string
amplitudes in \twoc\ dimensions in refs.~\citel{4d} and \citel{Corvi}.
However in the first paper it was not realised that the three-point
amplitude is non-zero, and in the second that the four-point amplitude is
zero.  The nature of the theory was therefore misunderstood.}.
The open string
sweeps out a world sheet that is a super-Riemann surface, but now with
boundaries.  For example, in the case of the super upper-half plane
which gives
the tree-level amplitudes of the theory, the boundary is given by $z = \zb
\equiv \sigma$, $\theta = \tb \equiv \theta$.  The action of
the string is the same as that of the closed string in eq\eq{wsaf}.  To
calculate with it, one also needs the boundary conditions for the fields,
which are given by
$\pa x = \pb
x |_{z=\zb}$ and $\psi_R=\psi_L \equiv \psi |_{z=\zb}$.

As in
the closed string, the spectrum of the open string is a single massless
scalar.  However, as is usually the case in open strings, we want to
append ``Chan-Paton'' group theory factors to the string amplitudes,
and so end up with a multiplet of scalars $\varphi^a$.
The superspace vertex operator
to emit these scalars
is the same as that of the closed scalars:
\beq
V_o = g \, e^{i(k \cdot \Xb + \kb \cdot X )} \comma
\eeq
but it has a different interpretation, since it is
inserted on the boundary of the super-Riemann surface. After integrating
out the fermionic coordinates one obtains
\beq
V_o^{int} = \int \dtt \, V_o =
\f{g}2 \, ( i k \cdot \pa_\sigma \xb -i \kb \cdot \pa_\sigma x -
4 k \cdot \psb \, \kb \cdot \psi )
\, e^{i(k \cdot \xb + \kb \cdot x )} \stop
\label{vomess}
\eeq
(The strange factors of 2 are because I am using
conventions appropriate to the closed string.)

The open-string three-point amplitude is then given by
\beq
\eqalign{
A_{ooo} &= \vev{V_o|_{\theta=0}(0) \cdot \int \dtt \, V_o(1)
\cdot V_o|_{\theta=0}(\infty)}  \cr
&= g \, c_{12} \times ( -i \, f^{abc} ) \comma
\label{a3o}
}
\eeq
where I have inserted the ``group-theory''
factor $-i f^{abc}$ by hand.  Without this factor the
amplitude would be totally antisymmetric with
respect to the three scalars, and so vanish. In the Chan-Paton \ansatz\
one would have $f^{abc}= \Tr \, ( \Lambda^a [ \Lambda^b \, , \,\Lambda^c ] ) $,
but since we have not yet established the principles for constructing \nq\
strings,
I shall here consider the most general
possible \ansatz{} consistent with principles to be given later.  Thus, at
this stage, $f^{abc}$ is a general unspecified totally antisymmetric
tensor.  (The notation may be a little suggestive, however.)
Note that, as usual\cite{openclosed}, the open amplitude
is the square root of the
closed-string amplitude of eq\eq{a3c}.  It is therefore the same as
that of the heterotic string, and can be derived from the
same field-theory action\cite{OVh}:
\beq
{\cal L}_{3o} = \int \dxf \, \left( \nfrac12 \, \pa^i \varphi^a \pb_{\ib}
\varphi^a -
i \,\f{g}3 \, f^{abc} \varphi^a \pa^i \varphi^b \pb_{\ib} \varphi^c \right)
+ O(\varphi^4) \stop
\label{cubic}
\eeq

The four-point amplitude of the string is given by
\beq
\eqalign{
A_{oooo} &\sim \int_0^1 \dx \vev{V_o|_{\theta=0}(0) \cdot
\int \dtt \, V_o(x) \cdot \int \dtt \, V_o(1)
\cdot V_o|_{\theta=0}(\infty)}  \cr
&= \f{g^2}{4} \,
F \; \f { \Gamma(1-2s) \, \Gamma(1-2t)} {\Gamma(2u) } \comma
\label{a4o}
}
\eeq
and, as in the closed string, vanishes because of the $F$ factor.
Because this amplitude vanishes, the usual unitarity
constraints\cite{meold} on the group-theory factors do not apply.
However, one can still find a constraint on them by calculating the four-point
function of the field theory action of eq\eq{cubic} (supplemented by
a quartic term ${\cal L}_{4o}$), and demanding that it
vanishes in agreement with the string result of eq\eq{a4o}.
Using some kinematical identities,
the field-theory result can be written as
\beq
\eqalign{
A_{oooo\,\hbox{\scriptsize FT}}
= -g^2 \bigg\{&
\f{c_{12}c_{34}}s \left ( f^{abx} f^{xcd} - f^{acx} f^{xbd}
	- f^{bcx} f^{xda} \right ) \cr
&+ u f^{bcx} f^{xda} + t f^{cax} f^{xbd} \bigg\} - V_{4o} \stop
\label{V4}
}
\eeq
Since $V_{4o}$ must be a local vertex,
the factor
in the parentheses must vanish for the amplitude to be zero.
This factor
is simply the Jacobi identity, so we see that the $f^{abc}$'s must be
the structure
constants of some group, as one might have been expected.  The group is
unspecified, and can be any semisimple group times a product of
$\uo$'s.  This is unlike the case of the bosonic and $N=1$ open strings,
where classically the group can only be $\son$, $\spn$ or
$\un$\cite{meold}\footnote{Of course, in the case of the superstring, we know
that the cancellation of anomalies uniquely picks out the group
$SO(32)$\cite{famous}.}.   We thus have an \ansatz\ for the group theory
that is more
general than  (but also even more ugly and \adhoc\/ than) that of Chan and
Paton\cite{CP}.
The resulting quartic interaction determined
by eq\eq{V4} is now\footnote{
Since the heterotic string also has a vanishing four-point amplitude,
this result is again in agreement with that of the pure Yang-Mills sector of
\citel{OVh}.  However, the heterotic
string also gets contribution from graphs with an
intermediate graviton, modifying ${\cal L}_{4o}$.
In an open theory such graphs have the topology
of an annulus, and are not part of the classical theory.}:
\beq
{\cal L}_{4o} = \int \dxf \, \left(- \f{g^2}6 \, f^{adx} f^{xbc} \,
\pa^i \varphi^a
\varphi^b \pb_{\ib} \varphi^c \varphi^d \right) \stop
\label{loooo}
\eeq

The action of eqs\eq{cubic} and (\ref{loooo}) does not appear to have any clear
meaning.  However, its resulting equation of motion can be written in a
nice compact way:  Defining the {\it anti-}hermitian matrix $\varphi$, the
equation of motion becomes
\beq
\pb_{\bar{i}} \left( e^{-2 i g \varphi} \pa^i e^{2 i g \varphi} \right) = 0
 \comma
\label{flat}
\eeq
which is known as Yang's equation\cite{Yang}.  As expected, since we again
have an equation with a name, it has a nice meaning and is indeed
related to self-dual Yang-Mills (SDYM).
This can be seen by noting that in a \kahler\ space the SDYM equation $F
= \tilde F$ breaks into three pieces.  Defining the holomorphic
$(2,0)$ form $\omega$ and the \kahler\ form $k$, one first has
\beq
\eqalign{
F_{ij} = 0 \quad (F \wedge \bar\omega = 0)
       &\; \iff \; A_i \equiv e^{-i g  \varphi} \pa_i e^{i g \varphi} \cr
F_{\ib \jb} = 0 \quad (F \wedge \omega = 0)
       &\; \iff \;
          \bar{A}_\ib \equiv e^{i g  \varphi} \pb_\ib e^{-i g \varphi}
\stop
\label{defs}
}
\eeq
(Here, choosing the $\varphi$'s in $A_i$ and $\bar{A}_\ib$ to be the same
means that we have fixed the gauge-invariance of the theory.)
The third self-duality equation is then
\beq
\eqalign{
F_i^\ib = 0 \quad (F \wedge k = 0)
       &\; \iff \; \left[ {\cal D}_i \, , \, \bar{\cal D}^\ib
        \right] = 0 \cr
       &\; \iff \;
        g^{i\jb} \pb_\jb
        \left( e^{-2 i g \varphi} \pa_i e^{2 i g \varphi} \right) = 0 \stop
\label{yang}
}
\eeq
Thus
the purely open string equation, eq\eq{flat}, describes SDYM in a
flat \twot-dimensional spacetime.    The
action giving
Yang's equation can be written order by order in $\varphi$, as we have
started to do in eqs\eq{a4o} and (\ref{loooo}).
Since the equation is a generalization of the
Wess-Zumino-Witten equation in two dimensions, it may not be surprising
that one can also write a more geometrical action in terms of coset
elements in a {\it five\/}-dimensional space\cite{NS}.

Summarizing, once the dust has settled, the equations of motion
of the string can again be
written in a completely Lorentz-invariant way:
\beq
F = \tilde F \p6!
\eeq

\vskip 0.2cm
{\it\noindent 2.4. Coupled open and closed \nq\ strings in \twoc-dimensions}
\vglue 0.1cm

We have seen that, as we expected, the open string describes SDYM.
However, we know
that all strings should contain gravity, so there should also be a
gravitational sector of the theory.  In the heterotic string the
gravitational sector indeed exists
and, as in the
\twor\ case, it changes the equations of motion of the string
from being exactly the self-duality of the Yang-Mills.
In open strings, gravity exists because open strings can always join
together to form closed ones.  Open theories thus always have closed
sectors, and one should consider interactions between the sectors.
Actually, not many calculations of this type have been carried out, at
least in the Polyakov formalism, so such calculations have
several relatively unfamiliar aspects to them.   For example, you should
bear in mind that in open string theories the tree-level equations of
the closed sector are found on the sphere, at genus 0, so the classical
gravitational equations of motion are the same
as those of the closed string.  However, the tree level of the open
sector comes from the disk, or upper-half plane (UHP), at ``genus
$1/2$'', so the very notion of the classical limit of the theory is not
so well defined.  For example, the closed and open-string couplings are
related by $\kappa \sim \sqrt \hbar g^2$.

I shall not go into the details of the mixed open-closed amplitude
calculations\cite{me}, but shall just give the results. First, all
amplitudes with only one open string and an arbitrary number of closed
ones vanish, because of the twist symmetry of the theory.  This is
fortunate, since the open sector has a group theory index and such
amplitudes imply a breaking of the group. The only
mixed three-point amplitude therefore
involves the scattering of two open strings
with one closed one on the UHP.  It is given by
\beq
\eqalign{
A_{ooc}
&\sim \int_{-\infty}^\infty \dx \vev{V_o|_{\theta=0}(x) \cdot \int
\dtt\dttb \, V_c(z=i)
\cdot V_o|_{\theta=0}(\infty)}  \cr
&= \vphantom{\int} \kappa \, \delta^{ab} c_{12}^2  \stop
\label{aooc}
}
\eeq
(Note that this amplitude involves an integration over the position of one
of the vertices, even though it is a three-point function!)
Since the amplitude gives the coupling of the gravitational sector to
the quadratic term of the open scalars, I have chosen
to append the group-theory factor $\delta^{ab}$, as in the kinetic
term.   This vertex is similar to that of the gravitational self-interaction
of eq\eq{a3c}, showing some kind of universality in the couplings of the
various fields to gravity.  The existence of the vertex means that one
has to add the interaction
\beq
{\cal L}_{ooc} = \int \dxf \, \left( 2\kappa \, \phi \, \pa\pb \varphi^a \wedge
\pa\pb \varphi^a \right)
\label{looc}
\eeq
to our open-closed action.

The first interesting four-point function involves the scattering of
three open strings with one closed one on the UHP.  It is given by
\beq
A_{oooc} = \nfrac{i}2 \kappa g f^{abc} \, F
         \, \f { \Gamma(s) \, \Gamma(t) \, \Gamma(u) }
                 { \Gamma(-s) \, \Gamma(-t) \, \Gamma(-u) }
                 \, \left( c_{12} t +c_{23} s \right)
\comma
\eeq
and it again vanishes on shell because of the $F$ factor.
Calculating the same amplitude in the field theory, one sees that one
has to add yet another term to the action:
\beq
{\cal L}_{oooc} = \int \dxf \, \left( - \nfrac43 \, i g \kappa \, f^{abc} \,
\pa \pb \phi \wedge \varphi^a
\pa \varphi^b \pb \varphi^c \right) \stop
\label{loooc}
\eeq
In principle, one should also consider the amplitude with two closed and two
open
scalars.  Since it is really messy to calculate, I shall assume that
it vanishes because of an $F$ factor, like all the other four-point
amplitudes.  This vanishing is found
from the field theory without having to add any new terms to the
Lagrangian.

There is one final type of amplitude that I have not yet discussed, because
it has a different structure and interpretation than all the other
amplitudes.  This is the scattering of three closed strings on the UHP and
on the projective plane $\rptwo$, which should be combined with the
UHP graph, since it is also of Euler number $1$.
Unlike graphs with open
vertices, whose tree-level amplitudes are defined on the UHP (since
$\rptwo$ has no boundaries, it does not appear in open amplitudes), the
closed three-point tree-level amplitude of the theory comes from the sphere.
These genus $1/2$ graphs thus give a
type of quantum corrections to the
theory.
Their amplitudes are difficult to calculate, but one can see that
they both have the form:
\beq
A_{ccc}' \propto \f{\kappa^3}{g^2} \, c_{12}^4 \comma
\label{half}
\eeq
with some finite coefficients.
Since the amplitudes contains no open-string
vertices, their group-theory factors are not fixed
and one can consider them
with arbitrary coefficients.  (In the Chan-Paton scheme
the overall factor would be proportional to $N-2$ for the group $\son$, \etc)
The interpretation of eq\eq{half} is unclear and, in the following, I will
generally choose the coefficients to make the
contribution vanish.

At this stage, we have enough of the action to see what the
full equations of motion of the theory should be.  The open-sector
equation is:
\beq
g_{i \jb} (\phi) \, \pb^{\bar{j}} \left( e^{-2 i g \varphi} \pa^i
e^{2 i g \varphi} \right) =
0
\; \iff \; F \wedge k = 0     \comma
\eeq
where $k(\phi)$ and $g(\phi)$ are now the full \Kahler\ form and metric,
defined in terms of $\phi$ as in
eq\eq{nk}.  This gives
the self-duality of the Yang-Mills field strength in
the background of
the curved
\Kahler{} space of the closed sector.
The closed equation of motion is now modified to
\beq
\det{g_{i \jb}} = -1 - \f{2\kappa^2}{g^2} \, \Tr \left( F_{i \jb} F^{i \jb}
 \right) \comma
\label{sd2}
\eeq
or
\beq
\pa \pb K \wedge \pa \pb K = 2 \omega \wedge \bar\omega
- \f{4\kappa^2}{g^2} \, \Tr \, \left( F \wedge F \right)
\stop
\label{sd1}
\eeq
Since the Ricci tensor is defined in terms of derivatives of
$\det g$ (eq\eq{ricci}), one sees that
the Ricci-flatness condition
is modified by a source term
from the open sector, in the same way that the Einstein tensor gets a
contribution from the matter stress tensor in usual gravity.
Recall that, since there is relation $\kappa \sim
\sqrt \hbar g^2$, the source term is some kind of quantum mechanical
correction to the equation of motion.  (If we had
included the quantum amplitude of eq\eq{half}, there would also have been a
gravitational source term on the right of eqs\eq{sd2} and (\ref{sd1}).)
The spacetime is thus no longer
Ricci flat, and is no longer self-dual.  I do not know of any
Lorentz-invariant way of writing this gravitational equation with the
source.

\vskip 0.2cm
{\it\noindent 2.5. Heterotic \nq\ strings in \twoc\ (?) dimensions}
\vglue 0.1cm

For completeness, I will now briefly discuss the heterotic
\nq\ strings\cite{OVh}.  The major problem with this string is
that the left-hand side of the string lives in \twot\ dimensions, while
the right-hand side lives in either $(25,1)$ or $(9,1)$ dimensions.  One
therefore needs to get rid of one of the LHS time coordinates.  This is
done, essentially, by compactifying the theory to either $(2,1)$ or
$(1,1)$ dimensions, somewhat messing up the geometrical interpretation of
the theories. As in the open case, the pure gauge sector of the theory is
described by Yang's equation for SDYM, now reduced to three (or two)
dimensions\cite{OVh}. Somewhat surprisingly, the resulting Yang-Mills
particles are actually tachyonic scalars! In addition, the theory also
contains massless vector-like particles in the gravitational sector, whose
couplings are rather poorly understood. As in
the $(1,1)$-dimensional case, the intermediate vector particles also induce an
$O(\alpha')$ modification to the equation of motion of the scalars, so
the gauge sector of the theory is not simply SDYM.  Since
our understanding of these theories is so confused,
I shall not say anything more about them in this talk.

\vskip 0.4cm
{\bf\noindent 3. Loops}
\vglue 0.2cm
{\it\noindent 3.1. The closed string partition function}
\vglue 0.1cm

Thus far, aside from the heterotic string, we have had a nice
interpretation of all of the amplitudes in the theory.  However once we
continue to loop amplitudes we are going to see all kinds of problems.  I
will give the results for several amplitudes here,
and discuss various possible ways around
the problems in
the conclusion of the talk.

The first loop calculation done in an \nq\ string---the partition
function of the closed string---was carried out by Mathur and
Mukhi\cite{part}.  Unlike other strings, this calculation
includes an integration over
the possible Wilson lines on the torus, which is equivalent to an
integration over the boundary conditions of the world-sheet fermions.
However, since each field in the theory is accompanied by a ghost field
of the same charge but different spin, and spin is irrelevant on the
torus, this means that {\it the partition function is independent of the
boundary conditions.\/}  After doing a careful integration
over all the zero modes in the theory, one obtains\cite{part}
\beq
Z_{\rm string} = \f1{4\pi} \int_{\cal M} \f{d \tau d \bar\tau} {\tau_2^{D/2}}
\label{pstr}
\comma
\eeq
where ${\cal M}$ is the usual ``keyhole'' of the moduli space of the
torus.
Recall that in the \twor-dimensional version of the theory one still has
all four scalars and only the $x$ zero-mode integration is changed, so
this result is valid both in \twor\ and \twoc\ dimensions.
The partition function is only modular invariant in the \twoc\ case, giving
the first evidence that this is where the theory should actually be
defined.

However, {\it the string partition is {\it not\/} that which one
would expect of a \twot-dimensional scalar, such as that of our Plebanski
lagrangian of eq\eq{pleb}!}\/  That would be given by
\beq
\eqalign{
Z_{part} &= \sum \, \nfrac12 \, \hbar \, \omega \cr
         &= \nfrac12 \; \Tr \, \log (p^2 +m^2)  \cr
         &= \nfrac12 \; \f1{(4\pi)^{D/2}} \int_0^\infty \f{d s}{s^{1+D/2}}
                                      e^{-s m^2}
\label{ppart} \comma
}
\eeq
where $s = \pi \alpha' \tau_2$ is the Schwinger parameter.  As usual, the
range of integration in the string case is different from that of the
particle\cite{Polchinski}.  What is really peculiar is that
the string result agrees with that of a massless particle in \twor\
and {\it not\/} \twoc\ dimensions.  Technically, this
is because the zero modes of the
extra $\uo$ ghosts of the string contribute an extra
factor of $\tau_2$, but this does not explain the physical discrepancy
between eqs\eq{pstr} and (\ref{ppart}).

The discrepancy also gives rise to another problem in \nq\ loop
amplitudes:  Instead of considering the string in a \twot-dimensional
Minkowski space, one can compactify one of the complex dimensions to
a complex torus.  This was done by Ooguri and Vafa\cite{OV}, using the
results of
Dixon \etal\/\cite{kapl}.  The partition function then
becomes
\beq
\eqalign{
Z_{\storus}
      &\sim \int_{\cal M} \f{d \tau d \bar\tau} {\tau_2} \cdot \sum \,
    q^{p_L^2/2} \bar{q}^{p_R^2/2} \cr
&\onArrow{\tau_2 \to \infty} \int^\infty
      \f{d \tau_2 } {\tau_2} \left( 1 + e^{-2 \pi M^2 \tau_2} + \cdots
\right) \comma
}
\eeq
and so develops a IR divergence as $\tau_2 \to \infty$.
Such divergences will turn out to plague the interpretation
of \nq\ loop amplitudes.

\vskip 0.2cm
{\it\noindent 3.2. The open string partition function}
\vglue 0.1cm

The open-string partition function shows many of the same problems:
In an open theory the partition function is found by calculating the path
integral on all the genus zero graphs: the torus, the Klein bottle, the
\Mobius{} band and the
annulus.  Defining the proper time $t$, the total partition function should be
given by
\beq
Z = \nfrac12 Z_{\storus} +
\f1{4\pi} \int_0^\infty \f{d t}{t^2} \left(\, \nfrac12 + \nfrac12 \,
c_{\sannulus}
- \nfrac12 \, c_{\smobius} \right)
\label{partfn}
\comma
\eeq
where the coefficients $c$ are group-theory factors.  (I have
put $c_{\sklein} = 1$, since I want the sum of the contributions of the
torus and
Klein bottle to give the one scalar of the closed sector.  The
factors of $1/2$ are due to the nonorientability of the theory, and
should be dropped if nonorientable graphs are discarded.)
In our general group-theory \ansatz, we do not
have any \apriori\ knowledge of the coefficients.  In order to get the spectrum
right, one needs the relation
\beq
c_{\sannulus} + c_{\smobius} = 2 \dim{G} \comma
\label{stupid}
\eeq
which is a new, not very obvious, constraint
on the \ansatz.  In the
Chan-Paton case this constraint is natural, since
$c_{\sannulus} = N^2$ and $c_{\smobius} = \pm N$ for
the groups $\son$ and $\spn$.

One can isolate the divergences of the partition function
by doing appropriate ``modular-like''
transformations on eq\eq{partfn}\cite{menew}.
The result is
\beq
Z = \nfrac12 Z_{\storus} +
\f1{16\pi} \left[ \, \int_0^1 \f{d q}{q} \left(2 + \nfrac12 \, c_{\sannulus}
\right) +
\int_{-1}^0 \f{d q}{q} \, 2 c_{\smobius} \, \right] \stop \label{part}
\eeq
This has
an IR divergence at $q=0$,
which in the Chan-Paton case can be regulated only for the
(very uninteresting) group $\sot$.
This is the
analogue of the groups $SO(32)$ for the superstring\cite{famous2} and
$SO(8192)$ for the bosonic string\cite{menew}.
However, since there are no known anomalies for the bosonic
and $N=2$ strings, the argument for these special groups is not very
compelling in these cases.

\vskip 0.2cm
{\it\noindent 3.3. One-loop three-point functions}
\vglue 0.1cm

So far we have seen problems in the calculation of partition functions.
One could argue that these are not very relevant physically, although this
would not be
the case for free energy calculations,
but actual scattering amplitude
calculations also show peculiar behaviour. This was first seen by Bonini,
Gava and Iengo\cite{italians} in the one-loop three-point scattering of
the closed string.  They found the result
\beq
A_3^{(1)} = 2 \, (2\pi)^6 g^3 c_{12}^6 \times \int_{\cal M} \f{d \tau d
\bar\tau}
{\tau_2^2} \, \f3{32\pi^6} \, {\sum_{n,m}}' \, \f{\tau_2^3}{|n+m\tau|^6}
\comma
\eeq
where $c_{12}$ is the usual kinematic factor of eq\eq{c12}.  This
amplitude presents several surprising features: Although it is a one-loop
amplitude, it is completely local!  (This is also the case for the similar
genus $1/2$ term of the open-closed theory in eq\eq{half}, but that is more
expected since it is not a true
quantum term.)  Also, the terms in the sum with
$m=0$ again give a nasty IR divergence.  Note that this
divergence can not be cured in the usual way:  Normally, one would
calculate the interference terms between this
graph and the tree-level graph in the
first-order correction to the three-point cross section.  The IR
divergence would then be expected to be canceled by the singular contribution
of
the tree-level four-point function.  However, in this theory the
four-point function
vanishes identically, so this cancellation can not work (even if one
knew how to define a cross-section in a spacetime with two times)!

A good thing about this amplitude is that, because of the simplicity
of the integrand, the integration over modular space
can (almost)
be carried out
explicitly\cite{menon}, using the techniques of Dixon \etal\/\cite{kapl}.
These allow one
to convert the sum over $m$ and $n$ in the amplitude
to a sum over
different copies of the moduli space, giving:
\beq
A_3^{(1)} = 2 \,(2\pi)^6 g^3 c_{12}^6 \times \int_0^\infty d \tau_2
                \; \f{\tau_2}{7!}
\label{loop}
\stop
\eeq
The ``almost'' in carrying out the calculation is that we are still stuck
with the IR divergence. Nevertheless, it is very interesting that the
string amplitude can be rewritten as an integral with a field-theory
Schwinger parameter, rather than as an integral over some complicated region in
moduli space.  This means that the amplitude can be directly compared to
the corresponding amplitude of the effective field theory, in the same way
as can the free-energy of string theories\cite{Polchinski}.  It
is a reasonable conjecture that, except for partition functions,
string theory amplitudes can always be rewritten in this way.
This means that common statements about ``strings naturally providing a UV
cut-off'', and so being intrinsically different from field theories are
not true, and string theories can really be studied in terms of their
effective field theories.

Unfortunately, when we attempt to compare this particular
amplitude to that of its field
theory we find the same problem that arose in the partition function
calculations. The amplitude calculated from the Plebanski action of
eq\eq{pleb} is
\beq
A_3^{(1)}\null_{FT} = \f1{(4\pi)^{D/2}} \; g^3 c_{12}^6 \times \int_0^\infty
d s \, \f{s^{2-D/2}} {7!} \stop
\eeq
This almost agrees with the string result (even up to the
$7!$, if not up to 2's and $\pi$'s!) except that, one again, the field
theory wants to live in \twor\ dimensions!

\vskip 0.4cm
{\bf\noindent 4. Conclusions}

\vglue 0.2cm
{\it\noindent 4.1. Tree level theory}
\vglue 0.1cm

Having told you all that I know about the amplitudes of the
\nq\ theories, I would now like to
stand back and to summarize the status of
the strings.  At the classical level, they are elegant and
well understood.  They are naturally defined in \twoc\ dimensions,
although one can consider them truncated to \twor\ dimensions.
This implies that they live in a
spacetime with signature \twot, and are theories with two times! Their
spectrum contains only massless scalar particles, so their space-time
string field
theories are simply field theories of the usual sort.  These
theories all turn out to have nice geometrical meanings:  In \twor\
dimensions, the closed string is trivial\cite{Mike}, while the open string
describes a sigma model theory, either on a group manifold\cite{adem} $G$,
or on the coset space\cite{OVh} $G_\CC/G$.  The heterotic string gives a
modified sigma model\cite{Mike}.  In \twoc\ dimensions, the closed theory
describes self-dual gravity (SDG) in a
\twot-dimensional
spacetime, with the
scalar of the string describing the \kahler\ potential of the
spacetime\cite{OV}.  The open string
describes self-dual Yang-Mills (SDYM) propagating in the \Kahler\
background of the closed sector\cite{me}, and its scalars parameterize the
gauge field as described by Yang\cite{Yang}.  The \nq\ open theory
can have any gauge group, since the usual unitarity
constraints\cite{meold} on the Chan-Paton factors no longer apply,
but the insertion of the group theory into the string is then very \adhoc.
The heterotic
string is less understood.  Its gauge sector also describes SDYM, with
a gauge group coming from the compactification of an internal
$24$-dimensional space, but the
theory must be compactified to $(2,1)$ dimensions\cite{OVh}.
This means that the scalars becomes
tachyonic!  In addition, the geometry of the gravity sector of the heterotic
string is poorly understood, and it in turn induces
interactions modifying the SDYM gauge structure.

The amplitudes of \nq\ strings are very unusual since, although the
strings only have massless particles, they must still have consistent dual
amplitudes.  This means that all the amplitudes of the theory---even
the loops!---are local in momenta.  In \twor\ dimensions, the (open and
heterotic) theories have nontrivial local four-point amplitudes; the next
possibly nonvanishing amplitude is the eight-point function, which has never
been calculated. In \twoc\ dimensions, all the theories have nonvanishing
three-point functions, which imply a ``physical'' decay process in the
\twot-dimensional spacetime.  All the four-point functions that have been
calculated vanish, and it is reasonable to conjecture that the
higher-point functions also vanish. This conjecture is related to the
folk-theorem that there is no classical scattering in SDYM\cite{belief},
and it has been suggested that one may be able to prove it using the
much-studied twistor formalism\cite{inst} for instantons.  However,
although there is a light-cone field theory calculation to support
it\cite{Parkes}, the conjecture remains unproven. The vanishing of these
amplitudes suggests that there is some kind of topological structure
to the \nq\ string theories, but it is not of the usual form.

\vskip 0.2cm
{\it\noindent 4.2. Loops and their problems}
\vglue 0.1cm

In contrast to the situation at the tree level, \nq\ loops are a bit of a
mess!  They present two problems:  First, they have nasty infra-red
divergences.  These are not intrinsically stringy, but can also be
seen in the effective field theory.  They may thus not seem to be so
surprising, since one is dealing with massless particles, but the
divergences cannot be cured in the usual ways.  In addition, when one tries
to compare the partition functions of the strings to those
of their effective field theories, the field-theory loop integrations must be
carried out in \twor\ dimensions to obtain the \twoc-dimensional result of
the strings\cite{part}!  A very nice feature of the \nq\ string is that,
because of the simplicity of its spectrum,
one can take a (closed three-point)
one-loop amplitude\cite{italians}, and rewrite it so that it
is explicitly an integration over a Schwinger parameter instead of
an integration over
moduli space\cite{menon}.  I conjecture that this
should be possible for all amplitudes in all string theories, except for
vacuum amplitudes. However, this particular \nq\ string amplitude also shows
both the IR divergence and the incompatibility of dimension with that
of its field theory.

\vskip 0.2cm
{\it\noindent 4.3. Lorentz invariance and the other spin structures}
\vglue 0.1cm

An important issue
that may have a bearing on the problems of the loop amplitudes,
is ``what is the Lorentz invariance of the \nq\ strings in
\twot\ dimensions?''.  As we have seen, the \nq\ theories are naturally
defined in a \kahler\ space, and so have a $\uoo$
invariance group. Since
the $\uoo$ Poincar\'e
group has no little group, this means that one can
not define the ``spin'' of the particles of the various sectors of the
theories.  However,
despite the \kahler\ nature of the theory (seen, for example, in the
ubiquitous $c_{12}$ factors in the amplitudes), the equations of motion of
the closed and open \nq\ strings can be written simply as $R=\tilde R$ and
$F=\tilde F$, respectively, and have a manifest $\sott$ Lorentz
invariance!  The reason for this is unclear.  While these equations of
motion are Lorentz invariant, it is well known that one can not write a
Lorentz invariant action for them without including new anti-self-dual
fields\cite{lorcov}, unless one uses nonlinear Lagrange multiplier
fields\cite{warlag}.  This makes it difficult to calculate
quantum corrections in the field theory
in a Lorentz invariant way.  There has been a recent
attempt to write a Lorentz-invariant action for SDYM in harmonic
space\cite{haction}, which seems very reasonable since the harmonic
approach is deeply related to twistors.  However, one can show that this
action does not give a correct description of the quantum
SDYM theory\cite{bunch}.

I should now like to amplify a bit more on the propositions of
Siegel concerning the spectrum and invariances of the \nq\
strings\cite{Warren,Warren2}.
These are rather controversial, and I shall not try to pronounce a final
verdict on them.  He first argues that the \nq\ strings
are the same as the ``$N=4$''
strings of \ad\cite{adem}, which have an $\sut$ world sheet
symmetry\cite{Warren}.  This would mean that the $x$ oscillators of the theory
can be taken to
be $\sott$ vectors, while the $\psi$'s would be $\sott$
spinors.
The constraint system of the $N=4$ theories is complicated to quantize, but
the equivalence immediately implies that \nq\ strings {\it must\/} be
completely Lorentz invariant.
This also has consequences for the non-NS sectors of the \nq\ string, which
I have generally ignored in the talk.  These sectors are normally regarded
as duplicates of the NS sector\cite{adem,OV}, but if
the theory is Lorentz invariant
the R sectors of the \nq\ strings must describe fermions\cite{Warren}.
Since no amplitudes of non-NS sectors have ever been calculated, there is as
yet no stringy evidence for or against this proposition.  The NS tree
amplitudes are unaffected by the existence of other sectors, and it is
hard to draw conclusions from the loop amplitudes, since they are so
confusing.

Siegel has also argued that the \nq\ strings should
actually be {\it maximally\/} spacetime supersymmetric\cite{Warren2}! This
conjecture is based on the fact that (without getting into nonlinear
Lagrange multipliers) one can write a Lorentz-invariant superspace action
for SDYM and SDG only if they have $N=4$ and $N=8$ spacetime
supersymmetries, respectively.  I feel that the arguments for this
conjecture can be evaded, since
there are supersymmetric theories which do not
have full-superspace actions\cite{n=4}, and
the ordinary type~II string does not have a usual Lorentz-invariant
action, since it has a self-dual field with no anti-dual partner.
If this proposition is nevertheless true,
it has very far-reaching consequences for the \nq\ theories:  First,
all the other fields of the extended supermultiplet must show up in
the theories.  For example, the other sectors of the
open string must give rise to {\it four\/} spinors
and {\it six\/} scalars,
in addition to the aforementioned anti-dual fields.
These numbers have to be included by hand, and appear to be somewhat \adhoc.
In the
case of the heterotic string, the gauge groups must also be far smaller
than those derived by Green\cite{Mike} and Ooguri and Vafa\cite{OVh}. In
the open string, it should not be possible for the open and closed sectors
to interact, since the two sectors have different $N=4$ and $N=8$
supersymmetries.  This contradicts to the various mixed amplitudes that
I have presented
here\cite{me}.  Somehow forcing these amplitudes to vanish would have the
good feature of removing the non-Lorentz invariant ``$1/2$-loop''
corrections
to the SDG equations
in eqs\eq{sd2} and (\ref{sd1}), but it also means that the SDYM of the
open string could live only in flat space!

The true loop
amplitudes of the various strings would also have to be changed:
supersymmetry means that the partition functions of all
the strings must vanish, with cancellations between the bosonic and
fermionic sectors of the theories.
This contradicts the standard result of ref.~\citel{part}, in which all
sectors contribute with the same phase.
In addition, the IR-divergent
non-Lorentz-invariant one loop closed-string amplitude of
eq\eq{loop}\cite{italians}, which gave us so much trouble, should also vanish.
In fact, the proposition of maximal supersymmetry solves all the loop
problems of the \nq\ theories rather dramatically: All loops must vanish
identically!

\vskip 0.2cm
{\it\noindent 4.4. Modifying string amplitudes?}
\vglue 0.1cm

All these results follow from
calculations in the (supposed) effective field theories of the \nq\
strings.  If they are to be relevant to the strings themselves,
one has to implement them at the string level.
I would like to conclude my talk by considering the possibility that the
\nq\ string amplitudes that I have talked about at such length
should be modified, and if so, how.

The mixed open-closed amplitudes
can only be made to vanish by declaring that they should not exist.  Since
I inserted group-theory factors by hand into these amplitudes, this is
possible to do.   It is, however, somewhat unsatisfying.  If one does
not wish to do this,
the gravity sector of the \nq\ open string could have at most an $N=4$
supersymmetry, as in the usual superstring, and not the maximal $N=8$
supersymmetry.

In the case of the true
one-loop amplitudes of eqs\eq{pstr} and (\ref{loop}),
one has more freedom to modify the string.
These amplitudes involve the adding together of different spin structures, all
of which give the same contribution.  Normally, one would add
them with fixed phases, but the only
convincing argument for this is that the final result should be modular
invariant and factorizable.  In the case of \nq\ strings all the sectors
are independently modular invariant, and unitarity and factorizability
are not understood in a spacetime with two times.
It thus may be reasonable to make the total amplitudes
vanish.  In fact, since the spacetime constraints on \nq\ strings are
so unclear, one could be even more extreme and drop
the possibility that the theories need be modular invariant: The
\twor-dimensional truncations of the theories could then be consistent.
While this may seem to be a remarkably ugly thing to do, the possibility can
not be ruled out by space-time arguments.

Finally, I should also point out that
all the calculations that I have presented here
are in the zero-instanton sector
of the theory.  The existence of the $\uo$ gauge field of the string means
that one should sum over world-sheet configurations with different instanton
number, possibly weighted with a $\theta$-term: $\theta \int \dzt F$.  The
necessity of such
calculations has been mentioned by various groups but they have never been
carried out.
One such calculation that we can do very
simply is the partition function on the torus:
Since the contribution of the charged
$\psi$ fields is canceled by that of the $(\beta,\gamma)$ ghosts, the
partition function is independent of the instanton number, giving:
\beq
\eqalign{
Z &\to Z_0 \times \sum_n e^{i n \theta} \cr
  &= 2 \pi \delta(\theta) \, Z_0  \stop
}
\eeq
This is a rather odd result (although it explains the strong-CP problem!),
and it may need to be modified---possibly to zero---as suggested
above.  The contribution of the sectors with nonvanishing instanton number
to scattering amplitudes should be further investigated.

As a final summing up: our investigations of the \nq\ strings have led to
many unexpected results.  The theories are still not
understood deeply, and we still need to find
the basic rules for constructing and calculating with these theories.
It should be interesting!

\newpage
{
\small
\def\baselinestretch{1}
\parskip=-4pt plus 2pt

}


\begin{thebibliography}{99}

\bibitem{ademonly}M.~Ademollo, L.~Brink, A.~D'Adda, R.~D'Auria,
        E.~Napolitano, S.~Sciuto, E.~Del Guido, P.~Di Vecchia, S.~Ferrara,
        F.~Gliozzi, R.~Musto and R.~Pettorini \pl62,76,105.
\bibitem{adem} M.~Ademollo, L.~Brink, A.~D'Adda, R.~D'Auria, E.~Napolitano,
        S.~Sciuto, E.~Del Guido, P.~Di Vecchia, S.~Ferrara, F.~Gliozzi,
        R.~Musto, R.~Pettorini and J.H.~Schwarz, \np111,76,77, \np114,76,297.
\bibitem{integ2} E.~Witten, \prl38,77,121;\\
         A.N.~Leznov and M.V.~Saveliev, \cmp74,80,111;\\
         L.~Mason and G.~Sparling, \pla137,89,29;\\
         I.~Bakas and D.A.~Depireux, \mpl6,91,399; \ijmp7,92,1767;
        \mpl6,91,1561, erratum-\ibid{} \noj{A6},91,2351.
\bibitem{conject} R.S.~Ward,
        \jou{Phil.\ Trans.\ Roy.\ Soc.\ Lond.},{A315},85,451;\\
         M.F.~Atiyah, unpublished;\\
         N.J.~Hitchin, \jou{Proc.\ Lond.\ Math.\ Soc.},5,87,59.
\bibitem{OV} H.~Ooguri and C.~Vafa, \mpl5,90,1389, \np361,91,469.
\bibitem{OVh} H.~Ooguri and C.~Vafa, \np367,91,83.
\bibitem{me} N.~Marcus, {\it ``The $N=2$ open string'',}\/ Tel-Aviv preprint
        {\bf TAUP--1929--91}, (HEP--TH/9207024),
        to appear in Nuclear Physics~B.
\bibitem{noncrit} I.~Antoniadis, C.~Bachas and C.~Kounnas, \pl242,90,185;\\
        J.~Distler, Z.~Hlousek and H.~Kawai, \ijmp5,90,391;\\
        H.~Lu, C.N.~Pope, X.J.~Wang and K.W.~Xu, \pl284,92,268.
\bibitem{BrinkS} L.~Brink and J.H.~Schwarz, \np121,77,285.
\bibitem{Frad} E.S.~Fradkin and A.A.~Tseytlin, \pl106,81,63.
\bibitem{4d}A.~D'Adda and F.~Lizzi, \pl191,87,85.
\bibitem{spect} A.~Schwimmer and N.~Seiberg, \pl184,87,191.
\bibitem{BRST} J.~Bie\'nkowska, \pl281,92,59.
\bibitem{Warren}  W.~Siegel, \prl69,92,1493.
\bibitem{Warren2}  W.~Siegel, {\it
        The $N=2(4)$ string is self-dual $N=4$ Yang-Mills,\/} ITP--SB--92--24,
        hep-th/9205075; {\it Self-dual $N=8$ supergravity as closed $N=2$
        $(N=4)$ strings,\/} ITP--SB--92--31, hep-th/9207043.
\bibitem{CP}  H.M.~Chan and J.~Paton, \np10,69,519.
\bibitem{Mike} M.B.~Green, \np293,87,593.
\bibitem{openclosed} H.~Kawai, D.C.~Lewellen and S.H.~Tye, \np269,86,1.
\bibitem{Plebanski}  J.F.~Plebanski, \jmp16,75,2395.
\bibitem{math} M.F.~Atiyah, N.J.~Hitchin and I.M.~Singer,
        \jou{Proc.\ R.\ Soc.\/ London},A362,78,425.
\bibitem{Corvi}  M.~Corvi, \pl231,89,240.
\bibitem{meold}  N.~Marcus and A.~Sagnotti, \pl119,82,97.
\bibitem{famous}  M.B.~Green and J.H.~Schwarz, \pl149,84,117.
\bibitem{Yang} C.N.~Yang \prl38,77,1377.
\bibitem{NS} V.P.~Nair and J.~Schiff, \pl246,90,423; \np371,92,329.
\bibitem{Polchinski}  J.~Polchinski, \cmp104,86,37.
\bibitem{part}  S.D.~Mathur and S.~Mukhi, \np302,88,130.
\bibitem{kapl}  L.J.~Dixon, V.~Kaplunovsky and J.~Louis, \np355,91,649.
\bibitem{menew}  N.~Marcus and A.~Sagnotti, \pl188,87,58.
\bibitem{famous2}  M.B.~Green and J.H.~Schwarz, \pl151,85,21.
\bibitem{italians}  M.~Bonini, E.~Gava and R.~Iengo, \mpl6,91,795.
\bibitem{menon}  N.~Marcus, unpublished.
\bibitem{belief} R.S.~Ward, {\it ``The twistor approach to differential
	equations'',\/} in {\it Quantum Gravity 2,\/} (Oxford 1980).
\bibitem{inst} R.S.~Ward, \pla61,77,81.\\
	M.F.~Atiyah and R.S.~Ward, \cmp55,77,117;\\
	E.F.~Corrigan, D.B.~Fairlie, R.G.~Yates and P.~Goddard, \cmp58,78,223.
\bibitem{Parkes}  A.~Parkes, \pl286,92,265.
\bibitem{lorcov} N.~Marcus and J.H.~Schwarz, \pl115,82,111.
\bibitem{warlag} W.~Siegel, \np238,84,307.
\bibitem{haction} S.~Kalitzin and E.~Sokatchev, \pl257,91,151.
\bibitem{bunch} N.~Marcus, Y.~Oz and S.~Yankielowicz, \np379,92,121.
\bibitem{n=4} W.~Siegel and M.~Ro\v{c}ek, \pl105,81,275.

\end{thebibliography}
\end{document}